\documentclass[aps,prb,twocolumn,showpacs,a4paper,balancelastpage,10pt]{revtex4-1}
\usepackage{epsfig}
\usepackage{graphicx}
\usepackage{color}
\usepackage{textcomp}
\usepackage{units}
\usepackage{ulem}

\input{tcilatex}
\begin{document}

\title{Tuning in magnetic modes in Tb(Co$_{x}$Ni$_{1-x}$)$_{2}$B$_{2}$C: \\
from longitudinal spin-density waves to simple ferromagnetism}
\author{M. ElMassalami,$^1$ H. Takeya,$^2$ B. Ouladdiaf,$^3$ R. Maia Filho,$%
^1$ A. M. Gomes,$^1$ T. Paiva,$^1$ and R. R. \surname{dos Santos}$^1$}
\affiliation{$^1$Instituto de Fisica, Universidade Federal do Rio de Janeiro, Caixa
Postal 68528, 21945-970 Rio de Janeiro, Brazil,\\
$^2$National Institute for Materials Science,1-2-1 Sengen, Tsukuba, Ibaraki
305-0047, Japan,\\
$^{3}$Institut Laue-Langevin, B.P. 156, F-38042 Grenoble Cedex 9, France}

\begin{abstract}
Neutron diffraction and thermodynamics techniques were used to probe the
evolution of the magnetic properties of \textrm{Tb(Co}$_{x}$\textrm{Ni}$%
_{1-x}$\textrm{)}$_{2}$\textrm{B}$_{2}$\textrm{C}. A{\ succession} of
magnetic modes was observed as $x$ is varied: the longitudinal modulated $%
\vec{k}=(0.55,0,0)$ state at $x=0$ is transformed into a collinear $\vec{k}=(%
\nicefrac{1}{2},0,\nicefrac{1}{2})$ antiferromagnetic state at $x=$ 0.2,
0.4; then into a transverse $c$-axis modulated $\vec{k}=(0,0,\nicefrac{1}{3}%
) $ mode at $x=$ 0.6, and finally into a simple ferromagnetic structure at $%
x=$ 0.8 and 1. Concomitantly, the low-temperature orthorhombic distortion of
the tetragonal unit cell at $x=0$ is reduced smoothly such that for $x\geq
0.4$ only a tetragonal unit cell is manifested. Though predicted
theoretically earlier, this is the first observation of the $\vec{k}=(0,0,%
\nicefrac{1}{3})$ mode in borocarbides; our findings of a{\ succession} of
magnetic modes upon increasing $x$ also find support from a recently
proposed theoretical model. The implication of these findings and their
interpretation on the magnetic structure of the $RM_{2}$\textrm{B}$_{2}$%
\textrm{C} \ series{\ are also discussed.}
\end{abstract}

\date{\today}
\maketitle

Rare-earth $4f$ moments at regular crystalline sites of intermetallic
matrices are subjected to a variety of competing interactions, such as the
Ruderman-Kittel-Kasuya-Yosida (RKKY), crystalline electric field,
magnetoelastic, and dipolar interactions.\cite%
{Coqblin-book,Jensen91-RareEarthMag} A particular class of such $4f$
intermetallics is the quaternary isomorphous borocarbides $RM_{2}$B$_{2}$C ($%
R$ is a rare earth or Y, and $M$ is a transition metal), which have been
found to exhibit coexistence between superconductivity and magnetism for a
judicious choice of $R$ and $M$.\cite%
{Nagarajan94-sup-(YNIBC),Cava94-sup-RNi2B2C,Muller01-interplay-review,Canfield-RNi2B2C-Hc2-review,03-Layer-Tc-borocarbides,Gupta06-Review-Borocarbides}
Apart from the interesting issue of coexistence (which highlights the
importance of electron interactions, with themselves as well as with the $4f$
moments), the magnetic properties of these materials pose a challenging
problem in their own right, especially when $R$ is magnetic and $M$ is
non-magnetic. For the interesting case of $M=$ Co or Ni, Table \ref%
{TabI-Mag-Tc} indicates that, for fixed $R$, the Co-based members exhibit
collinear and equal-amplitude ferromagnetic (FM)/antiferromagnetic (AFM)
structures;\cite{09-MS-RCo2B2C,09-MS-TbCo2B2C,09-MS-TmCo2B2C} by contrast,
the Ni-based members exhibit a variety of modulated structures, few
equal-amplitude and commensurate AFM structures, and an absence of FM modes.%
\cite{Lynn97-RNi2B2C-ND-mag-crys-structure}

Rhee \textit{et al.}\cite{Rhee95-generalized-susc} have calculated the
generalized susceptibility for borocarbides from band structures obtained
through the local-density approximation (LDA): for a fixed pair $R$ and $M$,
three incommensurate peaks were predicted near $\vec{k}_{1}=(0.6,0,0),$ $%
\vec{k}_{2}=(0,0,0.9)$ and $\vec{k}_{3}=(0,0,0.3)$. Though calculated for
the nonmagnetic LuNi$_{2}$B$_{2}$C, the results were expected to be valid
for all $R$, with the precise position and sharpness of each peak depending
on $R$ and $M$. Experimentally, both $\vec{k}_{1}$ and $\vec{k}_{2}$ modes
show up in HoNi$_{2}$B$_{2}$C and, moreover, $\vec{k}_{1}$ is evident in,
for example, $R=$ Er, Tb, Gd (Ref.%
\onlinecite{Lynn97-RNi2B2C-ND-mag-crys-structure}) but, so far, $\vec{k}_{3}$
has not been observed in borocarbides. It is recalled that this model does
not account for many modes [e.g. (0,0,1), ($\nicefrac{1}{2},0,\nicefrac{1}{2}
$) in $R$Ni$_{2}$B$_{2}$C and ($\nicefrac{1}{2},0,\nicefrac{1}{2}$), ($0,0,0$%
) $R$Co$_{2}$B$_{2}$C]; in Ref.\ \onlinecite{Rhee95-generalized-susc}, the
authors attributed this to the fact that their theory does not contemplate
any mechanism leading to an interplay between magnetism and
superconductivity. Then, a complementary theoretical approach contemplating
both magnetism and superconductivity, and from which their interplay can be
investigated, would be highly desirable. Bertussi \textit{et al.}\cite%
{Bertussi09-U-J-PhaseDiagram} proposed such a model, from which a succession
of magnetic modes exist even in the absence of superconductivity; see below{.%
}

\begin{table*}[t]
\caption{ Magnetic structures and transition temperatures, $T_{cr}$, of the
isomorphous $RM_{2}$B$_{2}$C ($R=$ Tm, Er, Ho, Dy, Tb; $M=$ Ni, Co) series.
TSDW (LSDW) denotes a transverse (longitudinal) modulated spin-density wave, 
$\vec{k}$ is the propagation wave vector, while $\vec{\protect\mu}$ is the
moment polarized along the easy axis. The magnetic properties of $R$Ni$_{2}$B%
$_{2}$C were taken from Refs.\ 
\onlinecite{Lynn97-RNi2B2C-ND-mag-crys-structure,Chang96-TmNi2B2C-mag-struct}%
, while those of $R$Co$_{2}$B$_{2}$C were taken from Refs.\ 
\onlinecite{09-MS-RCo2B2C,09-MS-TbCo2B2C,09-MS-TmCo2B2C}.}
\label{TabI-Mag-Tc}
\centering
\begin{tabular}{ccccccccccc}
\hline\hline
$R$ & \multicolumn{2}{c}{Tm} & \multicolumn{2}{c}{Er} & \multicolumn{2}{c}{Ho
} & \multicolumn{2}{c}{Dy} & \multicolumn{2}{c}{\textbf{Tb}} \\ \hline
$M$ & Co & Ni & Co & Ni & Co & Ni & Co & Ni & \textbf{Co} & \textbf{Ni} \\ 
\hline
$T_{cr}$ (K) & 0.8 & 1.53 & 4.0 & 6.8 & 5.4 & 5.0 & 8.0 & 10.6 & \textbf{6.3}
& \textbf{15.0} \\ 
structure & FM & TSDW & AFM & TSDW & FM & AFM & FM & AFM & \textbf{FM} & 
\textbf{LSDW} \\ 
$\overrightarrow{k}$ & (0,0,0) & (0.093,0.093,0) & ($\frac{1}{2},0,\frac{1}{2%
}$) & (0.553,0,0) & (0,0,0) & (0,0,1) & (0,0,0) & (0,0,1) & \textbf{(0,0,0)}
& \textbf{(0.555,0,0)} \\ 
$\left\vert \vec{\mu}\right\vert $ ($\mu_{B}$) & $\sim 1$ & 3.8 & 6.8(2) & 
7.2 & 7.2(2) & 8.6 & $>$5.2(2) & 8.5 & \textbf{7.6} & \textbf{7.8} \\ 
orientation & $\overrightarrow{c}$ & $\overrightarrow{c}$ & $\overrightarrow{%
b}$ & $\overrightarrow{b}$ & $\overrightarrow{a}+\overrightarrow{b}$ & $%
\overrightarrow{a}+\overrightarrow{b}$ & $\overrightarrow{a}+\overrightarrow{%
b}$ & $\overrightarrow{a}+\overrightarrow{b}$ & $\overrightarrow{a}$ & $%
\overrightarrow{a}$ \\ \hline\hline
\end{tabular}%
\end{table*}

This very particular case of the surge of various magnetic modes even in the
absence of superconductivity can best be tested in the \textrm{Tb(Co}$_{x}$%
\textrm{Ni}$_{1-x}$\textrm{)}$_{2}$\textrm{B}$_{2}$\textrm{C} series wherein
both end members are non-superconducting. With this in mind, here we report
on the mapping out of the magnetic modes of \textrm{Tb(Co}$_{x}$\textrm{Ni}$%
_{1-x}$\textrm{)}$_{2}$\textrm{B}$_{2}$\textrm{C} solid solutions. In view
of the markedly distinct magnetic structures of the Ni- and Co-based
compounds, it is certainly of interest to investigate in detail how{\ }the
magnetic (e.g., modes and moments) and structural properties develop as $M$
is changed (almost) continuously between these two limits. The choice of
this particular $R=$ Tb was dictated by the following features: (i) the
higher transition temperatures of the end members (in comparison with the
other pairs appearing in Table \ref{TabI-Mag-Tc}) allows for an
investigation over a wide temperature range; and (ii) the incommensurate
longitudinal spin density wave, LSDW, mode of \textrm{TbNi}$_{2}$\textrm{B}$%
_{2}$\textrm{C} is transformed into a commensurate FM state of \textrm{TbCo}$%
_{2}$\textrm{B}$_{2}$\textrm{C}, thus allowing, in principle, for several $%
\vec{k}$-vectors setting in. We will then be particularly interested in
elucidating whether this $M$-induced mode transformation is abrupt, or if
there are additional intermediate modes, in which case the determination of
how the $\vec{k}$-vectors are modified upon varying\ $M$ should highlight
the mechanisms at play. For completeness, we should mention that several
studies of the magnetic and electronic properties of $R(\mathrm{Co}_{x}%
\mathrm{Ni}_{1-x})_{2}$\textrm{B}$_{2}$\textrm{C} have been reported earlier
(see, e.g., Refs.\ {%
\onlinecite{Muller01-interplay-review,Gupta06-Review-Borocarbides}}, and
references therein).%
\begin{figure}[hptb]\centering
\includegraphics[
natheight=30.733cm, natwidth=21.3951cm, height=11.5345cm, width=8.0309cm]
{D:/Dropbox/TbNiCoB2C/NeutronDiffraction/graphics/TbCoxNiyB2C-ND-Fig1__1.pdf}%
\caption{(Color online) Neutron diffractograms of \textrm{Tb(Ni}$_{x}$%
\textrm{Co}$_{1-x}$\textrm{)}$_{2}$\textrm{B}$_{2}$C, measured at (a-d) $T=30
$ K, and (e-h) at $T=1.5$ K. \textit{Symbols:} measured intensities;\textit{%
\ vertical short bars:} Bragg positions of the nuclear and magnetic peaks; 
\textit{solid line:} Rietveld refined fit. The vertical dashed line
highlights theorthorhombic splitting of the tetragonal (3,0,5) peak into the
pairs: (3,0,5), (0,3,5). Inset: an expansion showing, for x=0.2, the single
peak at 30K (thin line) is orthrohmbic-split at 1.5 K into two peaks (thick
line). Space groups, positions, and occupations are given in text; thermal
parameters are the same as those reported by Lynn \textit{et al.}{
\onlinecite{Lynn97-RNi2B2C-ND-mag-crys-structure}}. $\vec{k}$ and $\vec{%
\protect\mu}$ are given in Table \protect\ref{TabII-Mag-CoxNi1-x}, while the
lattice parameters are shown in Fig.\ \protect\ref{Fig2-Crys-Tc}. The $R$
factors are in the range of 3-10.}\label{Fig1-ND}%
\end{figure}%

Polycrystalline samples of \textrm{Tb(Co}$_{x}$\textrm{Ni}$_{1-x}$\textrm{)}$%
_{2}$\textrm{B}$_{2}$\textrm{C} ($x=0,0.2,...,1$) with 99.5\% $^{11}$%
B-enriched were prepared by the conventional arc-melt method; all samples
were annealed\ for 20 hours at 1100$^{\circ }$C. Room-temperature
conventional x-ray diffraction analysis of polycrystalline samples (not
shown) indicated a single phase character for all compositions; the lattice
parameters, as obtained from the Rietveld analysis,\cite%
{Rodriguez-Carvajal93-Profile-Matching} are in excellent agreement with the
reported values.\cite%
{Siegrist94b-RNi2B2C,Lynn97-RNi2B2C-ND-mag-crys-structure,07-TbNi2B2C,09-MS-TbCo2B2C}
Powder neutron-diffractograms were collected at the high resolution powder
diffractometer D2B of the Institut Laue-Langevin (ILL), France $(\lambda =1.6%
{\mathrm{\mathring{A}}},T=1.5\mathrm{K}\,\mathrm{and}\,30\mathrm{K})$, and
were analyzed by the same Rietveld package. The diffractograms of the end
members were not measured since these have already been determined.\cite%
{Lynn97-RNi2B2C-ND-mag-crys-structure,Kawano08-TbNi2B2C-WF,09-MS-TbCo2B2C}
Magnetizations and susceptibilities ($2{\mathrm{K}}<T<20{\mathrm{K}}$ and $%
H\leq 90$ kOe) were measured on a Physical Properties Measurement System of
Quantum Design; specific heat{\ curves} ($1.8{\mathrm{K}}<T<40{\mathrm{K}}$
and $H=0,$ 30 kOe) were measured on a relaxation-type calorimeter using the
same environment as that used for the magnetization measurements. The
results obtained from these thermodynamic techniques (which will appear
elsewhere\cite{Tb(CoNi)2B2C-thermodynamic-2012}) provide independent
confirmation of the picture to be discussed below.

\begin{table*}[tbp]
\caption{ Magnetic properties of Tb(Co$_{x}$Ni$_{1-x}$)$_{2}$B$_{2}$C. $%
\protect\mu _{ND}$ is the zero-field moment as obtained from neutron
diffraction analysis, while $\protect\mu _{M}$ is the moment at 90 kOe as
obtained from magnetization isotherms at 2 K. Data for $x=0$ and $x=1$ were
taken from Refs.\ \onlinecite{Lynn97-RNi2B2C-ND-mag-crys-structure} and 
\onlinecite{09-MS-TbCo2B2C}, respectively. }
\label{TabII-Mag-CoxNi1-x}\centering
\par
\begin{tabular}{ccccccc}
\hline\hline
$x$ & 0 & 0.2 & 0.4 & 0.6 & 0.8 & 1.0 \\ \hline
$T_{cr}$ (K) & 15.0(2) & 12.0(2) & 8.3(3) & 8.8(3) & 7.4(2) & 6.6(2) \\ 
structure & LSDW & AFM & AFM & TSDW & FM & FM \\ 
$\vec{k}$ & ($0.55,0,0$) & ($\nicefrac{1}{2},0,\nicefrac{1}{2}$) & ($%
\nicefrac{1}{2},0,\nicefrac{1}{2}$) & ($0,0,\nicefrac{1}{3}$) & ($0,0,0$) & (%
$0,0,0$) \\ 
$\left\vert \vec{\mu}\right\vert _{\text{ND}}$ ($\mu_{B}$) & 7.8 & 7.6(1) & 
3.7(2) & 8.5(2) & 8.7(2) & 7.6 \\ 
$\left\vert \vec{\mu}\right\vert _{\text{M(90 kOe)}}$ ($\mu_{B}$) & 7.4(1) & 
4.1(1) & 7.6(2) & 7.7(1) & 7.6(1) & 7.2 \\ 
easy axis & $\vec{a}$ & $\vec{a}$ & $\vec{a}$ & $\vec{a}$ & $\vec{a}$ & $%
\vec{a}$ \\ \hline\hline
\end{tabular}%
\end{table*}

The 30\thinspace K nuclear diffractograms [Figs.\ \ref{Fig1-ND}(a)-(d)]
exhibit single-phase tetragonal structures ($I4/mmm$) with Tb, \textrm{Co}$%
_{x}$\textrm{Ni}$_{1-x}$, B, and C being at $2a,$ $4d,$ $4e,$ and $2b$
sites, respectively.\cite%
{Siegrist94b-RNi2B2C,Lynn97-RNi2B2C-ND-mag-crys-structure} On the other
hand, data at 1.5 K, Figs.\ \ref{Fig1-ND}(e)-(h), reveal a superposition of
magnetic and nuclear sub-patterns. Because of the well-known orthorhombic
distortion of \textrm{TbNi}$_{2}$\textrm{B}$_{2}$C $\left( \text{Refs.{%
\onlinecite{07-TbNi2B2C,Song99-Tb-magnetostriction,Song01-Tb-highResol-Mag-Xray}%
}}\right) $ no subtraction of the 30\thinspace K nuclear contribution was
attempted for the Ni-rich samples; rather, for the $x<0.4$ compositions, the
1.5\thinspace K nuclear subpatterns were analyzed assuming a
tetragonal-to-orthorhombic distorted unit cell [see inset of Fig. 1(e)].\
Applying this same analysis to the low-temperature $x\geq 0.4$
diffractograms yielded $a\approx b$; it is then clear that, for all $x\geq
0.4$, no structural distortion takes place. Accordingly, for these nuclear
diffractograms, the above mentioned $I4/mmm$ space group was used. The
whole-pattern fits are shown in Figs.\ \ref{Fig1-ND}(e)-\ref{Fig1-ND}(h),
and the (most important) fit parameters are shown in Table \ref%
{TabII-Mag-CoxNi1-x} and Figs.\ \ref{Fig2-Crys-Tc}(b)-\ref{Fig2-Crys-Tc}(d).%
\begin{figure}[ht]\centering
\includegraphics[
natheight=14.3725cm, natwidth=19.7235cm, height=5.8518cm, width=8.0331cm]
{D:/Dropbox/TbNiCoB2C/NeutronDiffraction/graphics/TbCoxNiyB2C-Param-Fig2__2.pdf}%
\caption{ (Color online) (a) The dependence with the Co fraction, $x$, of
the magnetic critical temperature $T_{cr}$, and of the lattice parameters:
(b) unit cell volume $V$, (c) lattice constants $c$, and (d) $a,b$. In
panels (b)-(d), stars and squares correspond, respectively, to fits to
tetragonal{\ (}30 K) and orthorhombic (1.5 K) unit cells; in panel (d),
filled and open symbols represent the $a$ and $b$ lattice constants. Dashed
lines are guides to the eye. Data for $x=0$ were taken from Refs.\ 
\onlinecite{07-TbNi2B2C,Song99-Tb-magnetostriction,Song01-Tb-highResol-Mag-Xray}%
, while those for $x=1$ were taken from Ref.\ \onlinecite{09-MS-TbCo2B2C}. }%
\label{Fig2-Crys-Tc}%
\end{figure}%

\begin{figure}[ht]\centering
\includegraphics[
natheight=8.5449cm, natwidth=17.2501cm, height=8.6679cm, width=8.0309cm]
{D:/Dropbox/TbNiCoB2C/NeutronDiffraction/graphics/TbCoxNiyB2C-MagStr-Fig3__3.pdf}%
\caption{The magnetic structures of \textrm{Tb(Co}$_{x}$\textrm{Ni}$_{1-x}$%
\textrm{)}$_{2}$\textrm{B}$_{2}$\textrm{C}. Neither the moment strength nor
the unit-cell dimensions were drawn to scale. Following Table \protect\ref%
{TabII-Mag-CoxNi1-x}, the Tb moments were taken to be polarized along the $%
\vec{a}$ axis.}\label{Fig3-MagStructure}%
\end{figure}%

The magnetic structures of Table \ref{TabII-Mag-CoxNi1-x} are visualized in
Fig.\ \ref{Fig3-MagStructure}. The moment orientation for each composition
was taken to be along the $a$ axis of the nuclear unit cell, based on the
reported features of the parent \textrm{Tb}$M_{2}$\textrm{B}$_{2}$C
compounds.\cite%
{07-TbNi2B2C,Song99-Tb-magnetostriction,Song01-Tb-highResol-Mag-Xray,Kawano08-TbNi2B2C-WF}
The incommensurate structure of the pure Ni sample ($x=0$) becomes, for $%
x=0.2$ and 0.4, a collinear AFM mode with $\vec{k}=$($\nicefrac{1}{2},0,%
\nicefrac{1}{2}$), \textit{i.e.}, AFM along the $a$ and $c$ axes, and FM
along the $b$ axis; this structure is different from the $a$-axis modulated
mode of \textrm{TbNi}$_{2}$\textrm{B}$_{2}$\textrm{C}, but similar to that
of ErCo$_{2}$B$_{2}$C,\cite{09-MS-RCo2B2C} and \textrm{NdNi}$_{2}$\textrm{B}$%
_{2}$\textrm{C}.\cite{Lynn97-RNi2B2C-ND-mag-crys-structure} One should note
that while the magnetic moment for $x=0.2$ is $\left\vert \vec{\mu}%
\right\vert =7.6\,\mu _{B}$ (thus very close to that for \textrm{TbNi}$_{2}$%
\textrm{B}$_{2}$\textrm{C}), for $x=0.4$ it is $\left\vert \vec{\mu}%
\right\vert =3.7(2)\,\mu _{B}$ which is, surprisingly, less than half of the
expected value. Such anomalous behavior is also evident in the
thermodynamical properties.\cite{Tb(CoNi)2B2C-thermodynamic-2012} In the
opposite limit of the Co-rich region, the $x=0.8$ sample displays a
commensurate $\vec{k}=(0,0,0)$ mode, just as for \textrm{TbCo}$_{2}$\textrm{B%
}$_{2}$\textrm{C};\cite{09-MS-TbCo2B2C} it should be noted that $|\vec{\mu}%
(x=0.8)|=8.7(2)\,\mu _{B}$, which is 14\% larger than $|\vec{\mu}%
(x=1)|=7.6\,\mu _{B}$.

The magnetic structure for $x=0.6$ is a transverse $c$-axis--modulated
spin-density--wave with $\vec{k}=(0,0,0.33\pm 0.01)\approx $ $(0,0,%
\nicefrac{1}{3})$: the FM planes are modulated, rotated, and stacked along
the $c$-axis with a period three times longer than that of the nuclear cell,
and with an amplitude of $10.8\,\mu _{B}$. At lower temperatures, this spin
desnity wave, SDW, will be squared-up due to the surge of higher, odd
Fourier harmonics;\cite{Lynn97-RNi2B2C-ND-mag-crys-structure} then, the
moment of this squared-up SDW will be, according to Fourier analysis, $(\pi
/4)\cdot 10.8\,\mu _{B}=8.5\,\mu _{B}$, in good agreement with the moment
found for the neighboring $x=0.8$ composition.

It should be noticed that the strength of the magnetic moment evolves
non-monotonically with $x$, with the lowest value being found within the
neighborhood of $x=0.4$. Furthermore, the observed propagation vectors, $%
\vec{k}=(0.55,0,0)$, ($\nicefrac{1}{2},0,\nicefrac{1}{2}$), ($0,0,%
\nicefrac{1}{3}$), and ($0,0,0$), form a subgroup of the main magnetic group
which, for borocarbides was predicted based on rigorous symmetry analyses of
representation theory.\cite{Wills03-IR-symmetry} Evidently, among{\ this
multitude} of $\vec{k}$ modes, the $\vec{k}$ = ($0,0,\nicefrac{1}{3}$) mode
is unique, since it has not been encountered in previous studies of either $%
R $\textrm{Co}$_{2}$\textrm{B}$_{2}$\textrm{C} (Refs.\ {%
\onlinecite{09-MS-RCo2B2C,09-MS-TbCo2B2C,09-MS-TmCo2B2C}}) or $R$\textrm{Ni}$%
_{2}$\textrm{B}$_{2}$\textrm{C} (Ref.\ %
\onlinecite{Lynn97-RNi2B2C-ND-mag-crys-structure}), though it is not
forbidden by representation theory.\cite{Wills03-IR-symmetry}

Our findings agree with the results from available theoretical approaches.
First, as mentioned before, the LDA calculations of Ref.\ %
\onlinecite{Rhee95-generalized-susc} predict three possible magnetic modes
in borocarbides, one of which, $\vec{k}_{3}$, has only been observed in the
present system. Second, a simple analysis suggests that the surge of a
succession of magnetic modes reported here may be attributed to a
competition between opposing tendencies of magnetic couplings. Indeed, the
oscillatory RKKY interaction between the local moments is mediated by the
conduction electrons, so that its spatial scale of oscillation is set by the
Fermi momentum $k_{F}$; then, due to electron count, $k_{F}$ would be
different in the Co-pure system and in the Ni-pure isomorph. Assuming a
continuously-varying effective $k_{F}$ upon `continuous' substitution of Ni
by Co, the associated variation in the RKKY coupling transforms the AFM mode
at the Ni-based limit into the FM mode at the Co-based limit through a
succession of intermediate modes.

Though this scenario is qualitatively consistent with our observations, a
step beyond this simplified picture is to contemplate superconductivity, and
its coexistence with magnetism, which in the context of the borocarbides is
crucial to reach a unified description. As mentioned in the Introduction
this task has been undertaken by Bertussi et al.,\cite%
{Bertussi09-U-J-PhaseDiagram} who proposed an \textit{effective} microscopic
model, in which the conduction electrons are subject to a pairing
interaction, say an attractive Hubbard-$U$ term, while they also mediate the
magnetic interaction between local moments via a Kondo-like coupling $J$.
The phase diagram obtained predicts a multitude of magnetic modes setting in
as $|J|$ increases, which can coexist (or not) with superconductivity,
depending on the relative strength of $|U|$ and $|J|$(see Ref.\ %
\onlinecite{Bertussi09-U-J-PhaseDiagram}); assuringly, such features are
consistent with the ones observed in borocarbides. In the present context of
non-superconducting Tb(Co$_{x}$Ni$_{1-x}$)$_{2}$B$_{2}$C, this model
corresponds to $U=0$, and predicts that a succession of magnetic modes are
stabilized as $|J|$ increases. According to the ground state phase diagram
of Ref.\ \onlinecite{Bertussi09-U-J-PhaseDiagram}, incommensurate
spin-density--waves (ISDW's), with a continuously varying{\ }$k$ vector,
exist for $|J|$ between 0 and some critical value, $J_{c1}$; the SDW becomes
commensurate (the AFM equivalent in one dimension, $k=\pi $) for $%
|J_{c1}|<|J|<|J_{c2}|$, and, finally, a FM state sets in for $|J|>|J_{c2}|$.
While the dependence of the effective parameter $|J|$ with Co concentration, 
$x$, cannot be extracted in a straightforward manner, the fact that the
model predicts the sequence observed with increasing $x$ can hardly be
regarded as fortuitous; the appearance of modes with continuously varying $k$
in Ref.\ \onlinecite{Bertussi09-U-J-PhaseDiagram}, instead of a single $\vec{%
k}_{3}$ mode may be attributed to the one-dimensional geometry of the
calculations.

This description, nonetheless, needs to be supplemented with other
ingredients so as to account for the above-mentioned anomalous behavior of
the $x=0.4$ sample.\ This indicates that the simple RKKY picture may not be
entirely applicable near this concentration, given that the magnitude of the
effective local moments is significantly reduced.\ Other theoretical
approaches, such as those of the authors of Refs.\ %
\onlinecite{Kalatsky98-clock-model,Amici-Thalmeier98-HoNi2B2C,Amici-Thalmeier-Fulde00-interplay}%
, have been used to describe the effects of an external field in the
magnetic phase diagram, but most likely would also need additional
ingredients to describe the behavior near $x=0.4$.

In summary, several experimental techniques have been used to study the
evolution of the magnetic properties of \textrm{Tb(Co}$_{x}$\textrm{Ni}$%
_{1-x}$\textrm{)}$_{2}$\textrm{B}$_{2}$\textrm{C}. The variation in $M=\text{%
Co}_{x}\text{Ni}_{1-x}$ modifies the electron count, and this, in turn,
introduces drastic variation in the magnetic structure leading to $\vec{k}$
modes compatible with symmetry requirements: $\vec{k}=(0.55,0,0)$ of the
Ni-based end member is transformed, successively, into $\vec{k}=(%
\nicefrac{1}{2},0,\nicefrac{1}{2})$ for $x=0.2,0.4$, $\vec{k}=(0,0,%
\nicefrac{1}{3})$ for $x=0.6$, and, finally, $\vec{k}=(0,0,0)$ for $x=0.8,1$%
. These modifications are accompanied by a lattice adjustment, indicative of
strong magnetoelastic forces. Magnetic anomalous behavior was observed in
the intermediate $x=0.4$ concentration. Finally, the confrontation of
available theoretical analyses with these results leads us to conclude that
the combined effect of electronic structure (i.e., the ensuing competition
between FM and AFM effective couplings) with magnetoelastic forces is
responsible for shaping both the lattice and magnetic properties of Tb(Co$%
_{x}$Ni$_{1-x}$)$_{2}$B$_{2}$C. This scenario can be easily generalized to
the wider case of $R$Ni$_{2}$B$_{2}$C, and $R$Co$_{2}$B$_{2}$C series.

\begin{acknowledgments}
Partial financial support from the Brazilian Agencies CNPq, CAPES, and
FAPERJ is gratefully acknowledged.
\end{acknowledgments}

\bibliographystyle{apsrev4-1}
\bibliography{borocarbides,crystalography,intermetallic,interplay-sup-mag,MagClassic,massalami,ND-RepAnalysis,notes,Sup-classic,supMagInterplay,To-Be-Published}

\begin{thebibliography}{26}%
\makeatletter
\providecommand \@ifxundefined [1]{%
 \@ifx{#1\undefined}
}%
\providecommand \@ifnum [1]{%
 \ifnum #1\expandafter \@firstoftwo
 \else \expandafter \@secondoftwo
 \fi
}%
\providecommand \@ifx [1]{%
 \ifx #1\expandafter \@firstoftwo
 \else \expandafter \@secondoftwo
 \fi
}%
\providecommand \natexlab [1]{#1}%
\providecommand \enquote  [1]{``#1''}%
\providecommand \bibnamefont  [1]{#1}%
\providecommand \bibfnamefont [1]{#1}%
\providecommand \citenamefont [1]{#1}%
\providecommand \href@noop [0]{\@secondoftwo}%
\providecommand \href [0]{\begingroup \@sanitize@url \@href}%
\providecommand \@href[1]{\@@startlink{#1}\@@href}%
\providecommand \@@href[1]{\endgroup#1\@@endlink}%
\providecommand \@sanitize@url [0]{\catcode `\\12\catcode `\$12\catcode
  `\&12\catcode `\#12\catcode `\^12\catcode `\_12\catcode `\%12\relax}%
\providecommand \@@startlink[1]{}%
\providecommand \@@endlink[0]{}%
\providecommand \url  [0]{\begingroup\@sanitize@url \@url }%
\providecommand \@url [1]{\endgroup\@href {#1}{\urlprefix }}%
\providecommand \urlprefix  [0]{URL }%
\providecommand \Eprint [0]{\href }%
\providecommand \doibase [0]{http://dx.doi.org/}%
\providecommand \selectlanguage [0]{\@gobble}%
\providecommand \bibinfo  [0]{\@secondoftwo}%
\providecommand \bibfield  [0]{\@secondoftwo}%
\providecommand \translation [1]{[#1]}%
\providecommand \BibitemOpen [0]{}%
\providecommand \bibitemStop [0]{}%
\providecommand \bibitemNoStop [0]{.\EOS\space}%
\providecommand \EOS [0]{\spacefactor3000\relax}%
\providecommand \BibitemShut  [1]{\csname bibitem#1\endcsname}%
\let\auto@bib@innerbib\@empty
\bibitem [{\citenamefont {Coqblin}(1977)}]{Coqblin-book}%
  \BibitemOpen
  \bibfield  {author} {\bibinfo {author} {\bibfnamefont {B.}~\bibnamefont
  {Coqblin}},\ }\href@noop {} {\emph {\bibinfo {title} {The Electronic
  Structure of Rare-Earth Metals $\newline$ and Aloys: The Magnetic Heavy
  Rare-Earth}}}\ (\bibinfo  {publisher} {Acdemic Press, New York},\ \bibinfo
  {year} {1977})\BibitemShut {NoStop}%
\bibitem [{\citenamefont {Jensen}\ and\ \citenamefont
  {Mackintosh}(1991)}]{Jensen91-RareEarthMag}%
  \BibitemOpen
  \bibfield  {author} {\bibinfo {author} {\bibfnamefont {J.}~\bibnamefont
  {Jensen}}\ and\ \bibinfo {author} {\bibfnamefont {A.~R.}\ \bibnamefont
  {Mackintosh}},\ }\href@noop {} {\emph {\bibinfo {title} {Rare Earth
  Magnetism: Structures And. Excitations}}}\ (\bibinfo  {publisher} {Clarendon
  Press, Oxfird},\ \bibinfo {year} {1991})\BibitemShut {NoStop}%
\bibitem [{\citenamefont {Nagarajan}\ \emph {et~al.}(1994)\citenamefont
  {Nagarajan}, \citenamefont {Mazumdar}, \citenamefont {Hossain}, \citenamefont
  {Dhar}, \citenamefont {Gopalakrishnan}, \citenamefont {Gupta}, \citenamefont
  {Godart}, \citenamefont {Padalia},\ and\ \citenamefont
  {Vijayaraghavan}}]{Nagarajan94-sup-(YNIBC)}%
  \BibitemOpen
  \bibfield  {author} {\bibinfo {author} {\bibfnamefont {R.}~\bibnamefont
  {Nagarajan}}, \bibinfo {author} {\bibfnamefont {C.}~\bibnamefont {Mazumdar}},
  \bibinfo {author} {\bibfnamefont {Z.}~\bibnamefont {Hossain}}, \bibinfo
  {author} {\bibfnamefont {S.~K.}\ \bibnamefont {Dhar}}, \bibinfo {author}
  {\bibfnamefont {K.~V.}\ \bibnamefont {Gopalakrishnan}}, \bibinfo {author}
  {\bibfnamefont {L.~C.}\ \bibnamefont {Gupta}}, \bibinfo {author}
  {\bibfnamefont {C.}~\bibnamefont {Godart}}, \bibinfo {author} {\bibfnamefont
  {B.~D.}\ \bibnamefont {Padalia}}, \ and\ \bibinfo {author} {\bibfnamefont
  {R.}~\bibnamefont {Vijayaraghavan}},\ }\href@noop {} {\bibfield  {journal}
  {\bibinfo  {journal} {Phys. Rev. Lett.}\ }\textbf {\bibinfo {volume} {72}},\
  \bibinfo {pages} {274} (\bibinfo {year} {1994})}\BibitemShut {NoStop}%
\bibitem [{\citenamefont {Cava}\ \emph {et~al.}(1994)\citenamefont {Cava},
  \citenamefont {Takagi}, \citenamefont {Zandbergen}, \citenamefont
  {Krajewski}, \citenamefont {F.~Peck~Jr.}, \citenamefont {Sigerist},
  \citenamefont {Batlogg}, \citenamefont {Dover}, \citenamefont {Felder},
  \citenamefont {Mizuhashi}, \citenamefont {Lee}, \citenamefont {Eisaki},\ and\
  \citenamefont {Uchida}}]{Cava94-sup-RNi2B2C}%
  \BibitemOpen
  \bibfield  {author} {\bibinfo {author} {\bibfnamefont {R.~J.}\ \bibnamefont
  {Cava}}, \bibinfo {author} {\bibfnamefont {H.}~\bibnamefont {Takagi}},
  \bibinfo {author} {\bibfnamefont {H.~W.}\ \bibnamefont {Zandbergen}},
  \bibinfo {author} {\bibfnamefont {J.~J.}\ \bibnamefont {Krajewski}}, \bibinfo
  {author} {\bibfnamefont {W.}~\bibnamefont {F.~Peck~Jr.}}, \bibinfo {author}
  {\bibfnamefont {T.}~\bibnamefont {Sigerist}}, \bibinfo {author}
  {\bibfnamefont {B.}~\bibnamefont {Batlogg}}, \bibinfo {author} {\bibfnamefont
  {R.~B.~V.}\ \bibnamefont {Dover}}, \bibinfo {author} {\bibfnamefont {R.~J.}\
  \bibnamefont {Felder}}, \bibinfo {author} {\bibfnamefont {K.}~\bibnamefont
  {Mizuhashi}}, \bibinfo {author} {\bibfnamefont {J.~O.}\ \bibnamefont {Lee}},
  \bibinfo {author} {\bibfnamefont {H.}~\bibnamefont {Eisaki}}, \ and\ \bibinfo
  {author} {\bibfnamefont {S.}~\bibnamefont {Uchida}},\ }\href@noop {}
  {\bibfield  {journal} {\bibinfo  {journal} {Nature}\ }\textbf {\bibinfo
  {volume} {367}},\ \bibinfo {pages} {254} (\bibinfo {year}
  {1994})}\BibitemShut {NoStop}%
\bibitem [{\citenamefont {M{\"u}ller}\ and\ \citenamefont
  {Narozhnyi}(2001)}]{Muller01-interplay-review}%
  \BibitemOpen
  \bibfield  {author} {\bibinfo {author} {\bibfnamefont {K.-H.}\ \bibnamefont
  {M{\"u}ller}}\ and\ \bibinfo {author} {\bibfnamefont {V.~N.}\ \bibnamefont
  {Narozhnyi}},\ }\bibfield  {booktitle} {\emph {\bibinfo {booktitle} {Rep.
  Prog. Phys.}},\ }\href@noop {} {\bibfield  {journal} {\bibinfo  {journal}
  {Rep. Prog. Phys.}\ }\textbf {\bibinfo {volume} {64}},\ \bibinfo {pages}
  {943} (\bibinfo {year} {2001})}\BibitemShut {NoStop}%
\bibitem [{\citenamefont {Canfield}\ \emph {et~al.}(1998)\citenamefont
  {Canfield}, \citenamefont {Gammel},\ and\ \citenamefont
  {Bishop}}]{Canfield-RNi2B2C-Hc2-review}%
  \BibitemOpen
  \bibfield  {author} {\bibinfo {author} {\bibfnamefont {P.~C.}\ \bibnamefont
  {Canfield}}, \bibinfo {author} {\bibfnamefont {P.~L.}\ \bibnamefont
  {Gammel}}, \ and\ \bibinfo {author} {\bibfnamefont {D.~J.}\ \bibnamefont
  {Bishop}},\ }\href@noop {} {\bibfield  {journal} {\bibinfo  {journal} {Phys.
  Today}\ }\textbf {\bibinfo {volume} {51}},\ \bibinfo {pages} {40} (\bibinfo
  {year} {1998})}\BibitemShut {NoStop}%
\bibitem [{\citenamefont {Paiva}\ \emph {et~al.}(2003)\citenamefont {Paiva},
  \citenamefont {El~Massalami},\ and\ \citenamefont {\surname {dos
  Santos}}}]{03-Layer-Tc-borocarbides}%
  \BibitemOpen
  \bibfield  {author} {\bibinfo {author} {\bibfnamefont {T.}~\bibnamefont
  {Paiva}}, \bibinfo {author} {\bibfnamefont {M.}~\bibnamefont {El~Massalami}},
  \ and\ \bibinfo {author} {\bibfnamefont {R.~R.}\ \bibnamefont {\surname {dos
  Santos}}},\ }\href@noop {} {\bibfield  {journal} {\bibinfo  {journal} {J.
  Phys.: Condens. Matter}\ }\textbf {\bibinfo {volume} {15}},\ \bibinfo {pages}
  {7917} (\bibinfo {year} {2003})}\BibitemShut {NoStop}%
\bibitem [{\citenamefont {Gupta}(2006)}]{Gupta06-Review-Borocarbides}%
  \BibitemOpen
  \bibfield  {author} {\bibinfo {author} {\bibfnamefont {L.~C.}\ \bibnamefont
  {Gupta}},\ }\href@noop {} {\bibfield  {journal} {\bibinfo  {journal} {Adv.
  Phys.}\ }\textbf {\bibinfo {volume} {55}},\ \bibinfo {pages} {691} (\bibinfo
  {year} {2006})}\BibitemShut {NoStop}%
\bibitem [{\citenamefont {ElMassalami}\ \emph
  {et~al.}(2009{\natexlab{a}})\citenamefont {ElMassalami}, \citenamefont
  {Moreno}, \citenamefont {Takeya}, \citenamefont {Ouladdiaf}, \citenamefont
  {Lynn},\ and\ \citenamefont {Freitas}}]{09-MS-RCo2B2C}%
  \BibitemOpen
  \bibfield  {author} {\bibinfo {author} {\bibfnamefont {M.}~\bibnamefont
  {ElMassalami}}, \bibinfo {author} {\bibfnamefont {R.}~\bibnamefont {Moreno}},
  \bibinfo {author} {\bibfnamefont {H.}~\bibnamefont {Takeya}}, \bibinfo
  {author} {\bibfnamefont {B.}~\bibnamefont {Ouladdiaf}}, \bibinfo {author}
  {\bibfnamefont {J.~W.}\ \bibnamefont {Lynn}}, \ and\ \bibinfo {author}
  {\bibfnamefont {R.~S.}\ \bibnamefont {Freitas}},\ }\href@noop {} {\bibfield
  {journal} {\bibinfo  {journal} {J. Phys: Condens Matter}\ }\textbf {\bibinfo
  {volume} {21}},\ \bibinfo {pages} {436006} (\bibinfo {year}
  {2009}{\natexlab{a}})}\BibitemShut {NoStop}%
\bibitem [{\citenamefont {ElMassalami}\ \emph
  {et~al.}(2009{\natexlab{b}})\citenamefont {ElMassalami}, \citenamefont
  {Moreno}, \citenamefont {Saeed}, \citenamefont {Chaves}, \citenamefont
  {Chaves}, \citenamefont {Rapp}, \citenamefont {Takeya}, \citenamefont
  {Ouladdiaf},\ and\ \citenamefont {Amara}}]{09-MS-TbCo2B2C}%
  \BibitemOpen
  \bibfield  {author} {\bibinfo {author} {\bibfnamefont {M.}~\bibnamefont
  {ElMassalami}}, \bibinfo {author} {\bibfnamefont {R.}~\bibnamefont {Moreno}},
  \bibinfo {author} {\bibfnamefont {R.~M.}\ \bibnamefont {Saeed}}, \bibinfo
  {author} {\bibfnamefont {F.~A.~B.}\ \bibnamefont {Chaves}}, \bibinfo {author}
  {\bibfnamefont {C.~M.}\ \bibnamefont {Chaves}}, \bibinfo {author}
  {\bibfnamefont {R.~E.}\ \bibnamefont {Rapp}}, \bibinfo {author}
  {\bibfnamefont {H.}~\bibnamefont {Takeya}}, \bibinfo {author} {\bibfnamefont
  {B.}~\bibnamefont {Ouladdiaf}}, \ and\ \bibinfo {author} {\bibfnamefont
  {M.}~\bibnamefont {Amara}},\ }\href@noop {} {\bibfield  {journal} {\bibinfo
  {journal} {J. Phys: Condens Matter}\ }\textbf {\bibinfo {volume} {21}},\
  \bibinfo {pages} {216006} (\bibinfo {year} {2009}{\natexlab{b}})}\BibitemShut
  {NoStop}%
\bibitem [{\citenamefont {ElMassalami}\ \emph
  {et~al.}(2009{\natexlab{c}})\citenamefont {ElMassalami}, \citenamefont
  {Rapp}, \citenamefont {Chaves}, \citenamefont {Moreno}, \citenamefont
  {Takeya}, \citenamefont {Ouladdiaf}, \citenamefont {Lynn}, \citenamefont
  {Huang}, \citenamefont {Freitas},\ and\ \citenamefont
  {Oliveria~Jr.}}]{09-MS-TmCo2B2C}%
  \BibitemOpen
  \bibfield  {author} {\bibinfo {author} {\bibfnamefont {M.}~\bibnamefont
  {ElMassalami}}, \bibinfo {author} {\bibfnamefont {R.~E.}\ \bibnamefont
  {Rapp}}, \bibinfo {author} {\bibfnamefont {F.~A.~B.}\ \bibnamefont {Chaves}},
  \bibinfo {author} {\bibfnamefont {R.}~\bibnamefont {Moreno}}, \bibinfo
  {author} {\bibfnamefont {H.}~\bibnamefont {Takeya}}, \bibinfo {author}
  {\bibfnamefont {B.}~\bibnamefont {Ouladdiaf}}, \bibinfo {author}
  {\bibfnamefont {J.~W.}\ \bibnamefont {Lynn}}, \bibinfo {author}
  {\bibfnamefont {Q.}~\bibnamefont {Huang}}, \bibinfo {author} {\bibfnamefont
  {R.~S.}\ \bibnamefont {Freitas}}, \ and\ \bibinfo {author} {\bibfnamefont
  {N.~F.}\ \bibnamefont {Oliveria~Jr.}},\ }\href@noop {} {\bibfield  {journal}
  {\bibinfo  {journal} {J. Phys.: Condens. Matter}\ }\textbf {\bibinfo {volume}
  {21}},\ \bibinfo {pages} {046007} (\bibinfo {year}
  {2009}{\natexlab{c}})}\BibitemShut {NoStop}%
\bibitem [{\citenamefont {Lynn}\ \emph {et~al.}(1997)\citenamefont {Lynn},
  \citenamefont {Skanthakumar}, \citenamefont {Huang}, \citenamefont {Sinha},
  \citenamefont {Hossain}, \citenamefont {Gupta}, \citenamefont {Nagarajan},\
  and\ \citenamefont {Godart}}]{Lynn97-RNi2B2C-ND-mag-crys-structure}%
  \BibitemOpen
  \bibfield  {author} {\bibinfo {author} {\bibfnamefont {J.~W.}\ \bibnamefont
  {Lynn}}, \bibinfo {author} {\bibfnamefont {S.}~\bibnamefont {Skanthakumar}},
  \bibinfo {author} {\bibfnamefont {Q.}~\bibnamefont {Huang}}, \bibinfo
  {author} {\bibfnamefont {S.~K.}\ \bibnamefont {Sinha}}, \bibinfo {author}
  {\bibfnamefont {Z.}~\bibnamefont {Hossain}}, \bibinfo {author} {\bibfnamefont
  {L.~C.}\ \bibnamefont {Gupta}}, \bibinfo {author} {\bibfnamefont
  {R.}~\bibnamefont {Nagarajan}}, \ and\ \bibinfo {author} {\bibfnamefont
  {C.}~\bibnamefont {Godart}},\ }\href@noop {} {\bibfield  {journal} {\bibinfo
  {journal} {Phys. Rev. B}\ }\textbf {\bibinfo {volume} {55}},\ \bibinfo
  {pages} {6584} (\bibinfo {year} {1997})}\BibitemShut {NoStop}%
\bibitem [{\citenamefont {Rhee}\ \emph {et~al.}(1995)\citenamefont {Rhee},
  \citenamefont {Wang},\ and\ \citenamefont
  {Harmon}}]{Rhee95-generalized-susc}%
  \BibitemOpen
  \bibfield  {author} {\bibinfo {author} {\bibfnamefont {J.~Y.}\ \bibnamefont
  {Rhee}}, \bibinfo {author} {\bibfnamefont {X.}~\bibnamefont {Wang}}, \ and\
  \bibinfo {author} {\bibfnamefont {B.~N.}\ \bibnamefont {Harmon}},\
  }\href@noop {} {\bibfield  {journal} {\bibinfo  {journal} {Phys. Rev. B}\
  }\textbf {\bibinfo {volume} {51}},\ \bibinfo {pages} {15585} (\bibinfo {year}
  {1995})}\BibitemShut {NoStop}%
\bibitem [{\citenamefont {Bertussi}\ \emph {et~al.}(2009)\citenamefont
  {Bertussi}, \citenamefont {Malvezzi}, \citenamefont {Paiva},\ and\
  \citenamefont {Santos}}]{Bertussi09-U-J-PhaseDiagram}%
  \BibitemOpen
  \bibfield  {author} {\bibinfo {author} {\bibfnamefont {P.~R.}\ \bibnamefont
  {Bertussi}}, \bibinfo {author} {\bibfnamefont {A.~L.}\ \bibnamefont
  {Malvezzi}}, \bibinfo {author} {\bibfnamefont {T.}~\bibnamefont {Paiva}}, \
  and\ \bibinfo {author} {\bibfnamefont {R.~R.~D.}\ \bibnamefont {Santos}},\
  }\href@noop {} {\bibfield  {journal} {\bibinfo  {journal} {Phys. Rev. B}\
  }\textbf {\bibinfo {volume} {79}},\ \bibinfo {pages} {220513} (\bibinfo
  {year} {2009})}\BibitemShut {NoStop}%
\bibitem [{\citenamefont {Chang}\ \emph {et~al.}(1996)\citenamefont {Chang},
  \citenamefont {Tomy}, \citenamefont {Paul},\ and\ \citenamefont
  {Ritter}}]{Chang96-TmNi2B2C-mag-struct}%
  \BibitemOpen
  \bibfield  {author} {\bibinfo {author} {\bibfnamefont {L.~J.}\ \bibnamefont
  {Chang}}, \bibinfo {author} {\bibfnamefont {C.~V.}\ \bibnamefont {Tomy}},
  \bibinfo {author} {\bibfnamefont {D.~M.}\ \bibnamefont {Paul}}, \ and\
  \bibinfo {author} {\bibfnamefont {C.}~\bibnamefont {Ritter}},\ }\href@noop {}
  {\bibfield  {journal} {\bibinfo  {journal} {Phys. Rev. B}\ }\textbf {\bibinfo
  {volume} {54}},\ \bibinfo {pages} {9031} (\bibinfo {year}
  {1996})}\BibitemShut {NoStop}%
\bibitem [{\citenamefont
  {Rodr{\'{\i}}guez-Carvajal}(1993)}]{Rodriguez-Carvajal93-Profile-Matching}%
  \BibitemOpen
  \bibfield  {author} {\bibinfo {author} {\bibfnamefont {J.}~\bibnamefont
  {Rodr{\'{\i}}guez-Carvajal}},\ }\href@noop {} {\bibfield  {journal} {\bibinfo
   {journal} {Physica B}\ }\textbf {\bibinfo {volume} {192}},\ \bibinfo {pages}
  {55} (\bibinfo {year} {1993})}\BibitemShut {NoStop}%
\bibitem [{\citenamefont {Siegrist}\ \emph {et~al.}(1994)\citenamefont
  {Siegrist}, \citenamefont {Cava}, \citenamefont {Krajewski},\ and\
  \citenamefont {Peck}}]{Siegrist94b-RNi2B2C}%
  \BibitemOpen
  \bibfield  {author} {\bibinfo {author} {\bibfnamefont {T.}~\bibnamefont
  {Siegrist}}, \bibinfo {author} {\bibfnamefont {R.}~\bibnamefont {Cava}},
  \bibinfo {author} {\bibfnamefont {J.~J.}\ \bibnamefont {Krajewski}}, \ and\
  \bibinfo {author} {\bibfnamefont {W.~F.}\ \bibnamefont {Peck}},\ }\href@noop
  {} {\bibfield  {journal} {\bibinfo  {journal} {J. Alloys Compd.}\ }\textbf
  {\bibinfo {volume} {216}},\ \bibinfo {pages} {135} (\bibinfo {year}
  {1994})}\BibitemShut {NoStop}%
\bibitem [{\citenamefont {ElMassalami}\ \emph {et~al.}(2007)\citenamefont
  {ElMassalami}, \citenamefont {Amara}, \citenamefont {Galera}, \citenamefont
  {Schmitt},\ and\ \citenamefont {Takeya}}]{07-TbNi2B2C}%
  \BibitemOpen
  \bibfield  {author} {\bibinfo {author} {\bibfnamefont {M.}~\bibnamefont
  {ElMassalami}}, \bibinfo {author} {\bibfnamefont {M.}~\bibnamefont {Amara}},
  \bibinfo {author} {\bibfnamefont {R.-M.}\ \bibnamefont {Galera}}, \bibinfo
  {author} {\bibfnamefont {D.}~\bibnamefont {Schmitt}}, \ and\ \bibinfo
  {author} {\bibfnamefont {H.}~\bibnamefont {Takeya}},\ }\href@noop {}
  {\bibfield  {journal} {\bibinfo  {journal} {Phys. Rev. B}\ }\textbf {\bibinfo
  {volume} {76}},\ \bibinfo {pages} {104410} (\bibinfo {year}
  {2007})}\BibitemShut {NoStop}%
\bibitem [{\citenamefont {Kawano-Furukawa}\ \emph {et~al.}(2008)\citenamefont
  {Kawano-Furukawa}, \citenamefont {Tsukagoshi}, \citenamefont {Nagata},
  \citenamefont {Kobayashi}, \citenamefont {Yoshizawa},\ and\ \citenamefont
  {Takeya}}]{Kawano08-TbNi2B2C-WF}%
  \BibitemOpen
  \bibfield  {author} {\bibinfo {author} {\bibfnamefont {H.}~\bibnamefont
  {Kawano-Furukawa}}, \bibinfo {author} {\bibfnamefont {H.}~\bibnamefont
  {Tsukagoshi}}, \bibinfo {author} {\bibfnamefont {T.}~\bibnamefont {Nagata}},
  \bibinfo {author} {\bibfnamefont {C.}~\bibnamefont {Kobayashi}}, \bibinfo
  {author} {\bibfnamefont {H.}~\bibnamefont {Yoshizawa}}, \ and\ \bibinfo
  {author} {\bibfnamefont {H.}~\bibnamefont {Takeya}},\ }\href@noop {}
  {\bibfield  {journal} {\bibinfo  {journal} {Phys. Rev. B}\ }\textbf {\bibinfo
  {volume} {77}},\ \bibinfo {pages} {144426} (\bibinfo {year}
  {2008})}\BibitemShut {NoStop}%
\bibitem [{\citenamefont {\surname{ElMassalami {\it et
  al.}}}(2012)}]{Tb(CoNi)2B2C-thermodynamic-2012}%
  \BibitemOpen
  \bibfield  {author} {\bibinfo {author} {\bibfnamefont {M.}~\bibnamefont
  {\surname{ElMassalami {\it et al.}}}},\ }\href@noop {} {\  (\bibinfo {year}
  {2012})},\ \bibinfo {note} {(unpublished)}\BibitemShut {NoStop}%
\bibitem [{\citenamefont {Song}\ \emph {et~al.}(1999)\citenamefont {Song},
  \citenamefont {Islam}, \citenamefont {Lottermoser}, \citenamefont {Goldman},
  \citenamefont {Canfield},\ and\ \citenamefont
  {Detlefs}}]{Song99-Tb-magnetostriction}%
  \BibitemOpen
  \bibfield  {author} {\bibinfo {author} {\bibfnamefont {C.}~\bibnamefont
  {Song}}, \bibinfo {author} {\bibfnamefont {Z.}~\bibnamefont {Islam}},
  \bibinfo {author} {\bibfnamefont {L.}~\bibnamefont {Lottermoser}}, \bibinfo
  {author} {\bibfnamefont {A.~I.}\ \bibnamefont {Goldman}}, \bibinfo {author}
  {\bibfnamefont {P.~C.}\ \bibnamefont {Canfield}}, \ and\ \bibinfo {author}
  {\bibfnamefont {C.}~\bibnamefont {Detlefs}},\ }\href@noop {} {\bibfield
  {journal} {\bibinfo  {journal} {Phys. Rev. B}\ }\textbf {\bibinfo {volume}
  {60}},\ \bibinfo {pages} {6223} (\bibinfo {year} {1999})}\BibitemShut
  {NoStop}%
\bibitem [{\citenamefont {Song}\ \emph {et~al.}(2001)\citenamefont {Song},
  \citenamefont {Wermeille}, \citenamefont {Goldman}, \citenamefont {Canfield},
  \citenamefont {Rhee},\ and\ \citenamefont
  {Harmon}}]{Song01-Tb-highResol-Mag-Xray}%
  \BibitemOpen
  \bibfield  {author} {\bibinfo {author} {\bibfnamefont {C.}~\bibnamefont
  {Song}}, \bibinfo {author} {\bibfnamefont {D.}~\bibnamefont {Wermeille}},
  \bibinfo {author} {\bibfnamefont {A.~I.}\ \bibnamefont {Goldman}}, \bibinfo
  {author} {\bibfnamefont {P.~C.}\ \bibnamefont {Canfield}}, \bibinfo {author}
  {\bibfnamefont {J.~Y.}\ \bibnamefont {Rhee}}, \ and\ \bibinfo {author}
  {\bibfnamefont {B.~N.}\ \bibnamefont {Harmon}},\ }\href@noop {} {\bibfield
  {journal} {\bibinfo  {journal} {Phys. Rev. B}\ }\textbf {\bibinfo {volume}
  {63}},\ \bibinfo {pages} {104 507} (\bibinfo {year} {2001})}\BibitemShut
  {NoStop}%
\bibitem [{\citenamefont {Wills}\ \emph {et~al.}(2003)\citenamefont {Wills},
  \citenamefont {Detlefs},\ and\ \citenamefont
  {Canfield}}]{Wills03-IR-symmetry}%
  \BibitemOpen
  \bibfield  {author} {\bibinfo {author} {\bibfnamefont {A.~S.}\ \bibnamefont
  {Wills}}, \bibinfo {author} {\bibfnamefont {C.}~\bibnamefont {Detlefs}}, \
  and\ \bibinfo {author} {\bibfnamefont {P.~C.}\ \bibnamefont {Canfield}},\
  }\href@noop {} {\bibfield  {journal} {\bibinfo  {journal} {Phil. Mag.}\
  }\textbf {\bibinfo {volume} {83}},\ \bibinfo {pages} {1227} (\bibinfo {year}
  {2003})}\BibitemShut {NoStop}%
\bibitem [{\citenamefont {Kalatsky}\ and\ \citenamefont
  {Pokrovsky}(1998)}]{Kalatsky98-clock-model}%
  \BibitemOpen
  \bibfield  {author} {\bibinfo {author} {\bibfnamefont {V.~A.}\ \bibnamefont
  {Kalatsky}}\ and\ \bibinfo {author} {\bibfnamefont {V.~L.}\ \bibnamefont
  {Pokrovsky}},\ }\href@noop {} {\bibfield  {journal} {\bibinfo  {journal}
  {Phys. Rev. B}\ }\textbf {\bibinfo {volume} {57}},\ \bibinfo {pages} {5485}
  (\bibinfo {year} {1998})}\BibitemShut {NoStop}%
\bibitem [{\citenamefont {Amici}\ and\ \citenamefont
  {Thalmeier}(1998)}]{Amici-Thalmeier98-HoNi2B2C}%
  \BibitemOpen
  \bibfield  {author} {\bibinfo {author} {\bibfnamefont {A.}~\bibnamefont
  {Amici}}\ and\ \bibinfo {author} {\bibfnamefont {P.}~\bibnamefont
  {Thalmeier}},\ }\href@noop {} {\bibfield  {journal} {\bibinfo  {journal}
  {Phys. Rev. B}\ }\textbf {\bibinfo {volume} {57}},\ \bibinfo {pages} {10 684}
  (\bibinfo {year} {1998})}\BibitemShut {NoStop}%
\bibitem [{\citenamefont {Amici}\ \emph {et~al.}(2000)\citenamefont {Amici},
  \citenamefont {Thalmeier},\ and\ \citenamefont
  {Fulde}}]{Amici-Thalmeier-Fulde00-interplay}%
  \BibitemOpen
  \bibfield  {author} {\bibinfo {author} {\bibfnamefont {A.}~\bibnamefont
  {Amici}}, \bibinfo {author} {\bibfnamefont {P.}~\bibnamefont {Thalmeier}}, \
  and\ \bibinfo {author} {\bibfnamefont {P.}~\bibnamefont {Fulde}},\
  }\href@noop {} {\bibfield  {journal} {\bibinfo  {journal} {Phys. Rev. Lett.}\
  }\textbf {\bibinfo {volume} {84}},\ \bibinfo {pages} {1800} (\bibinfo {year}
  {2000})}\BibitemShut {NoStop}%
\end{thebibliography}%

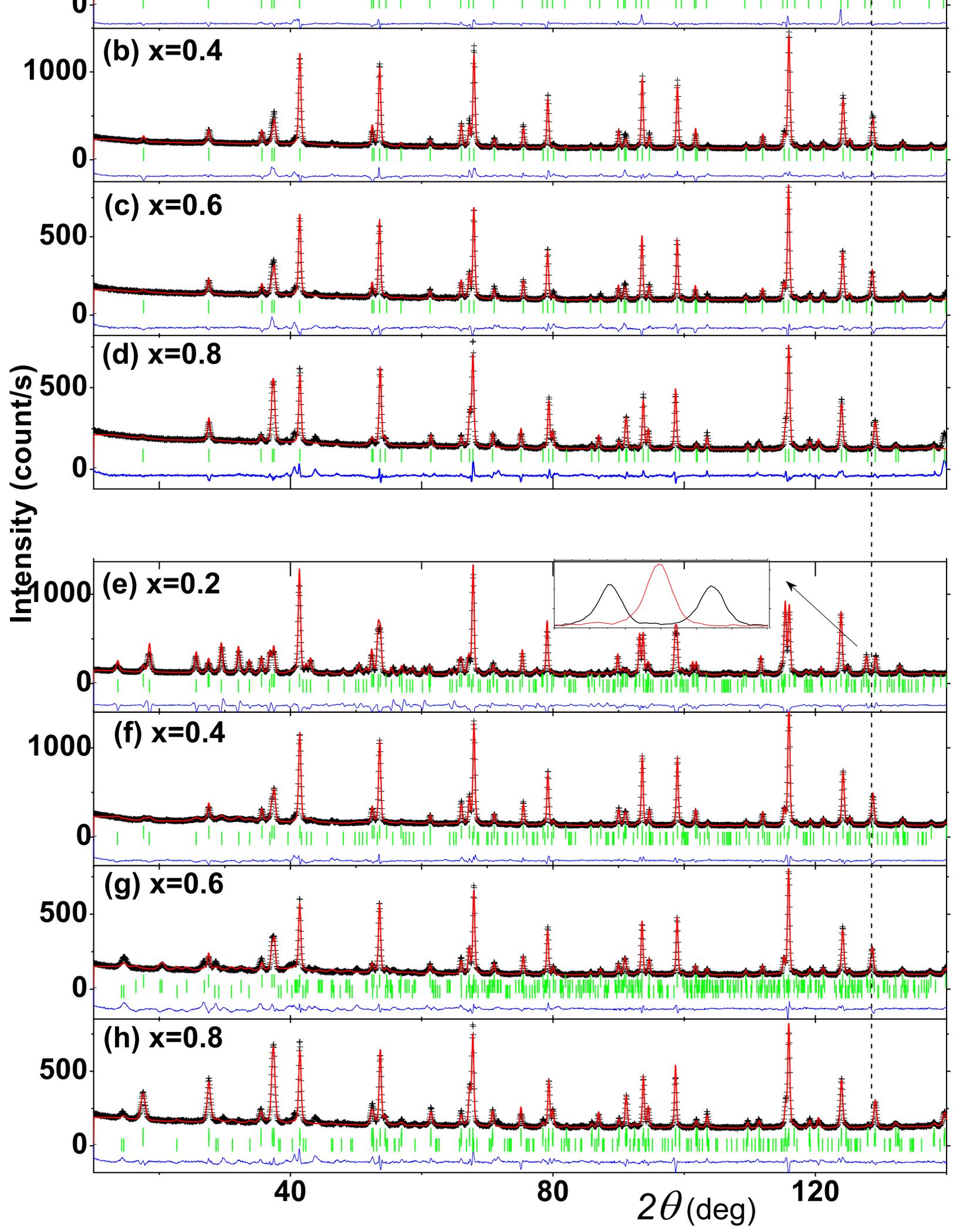
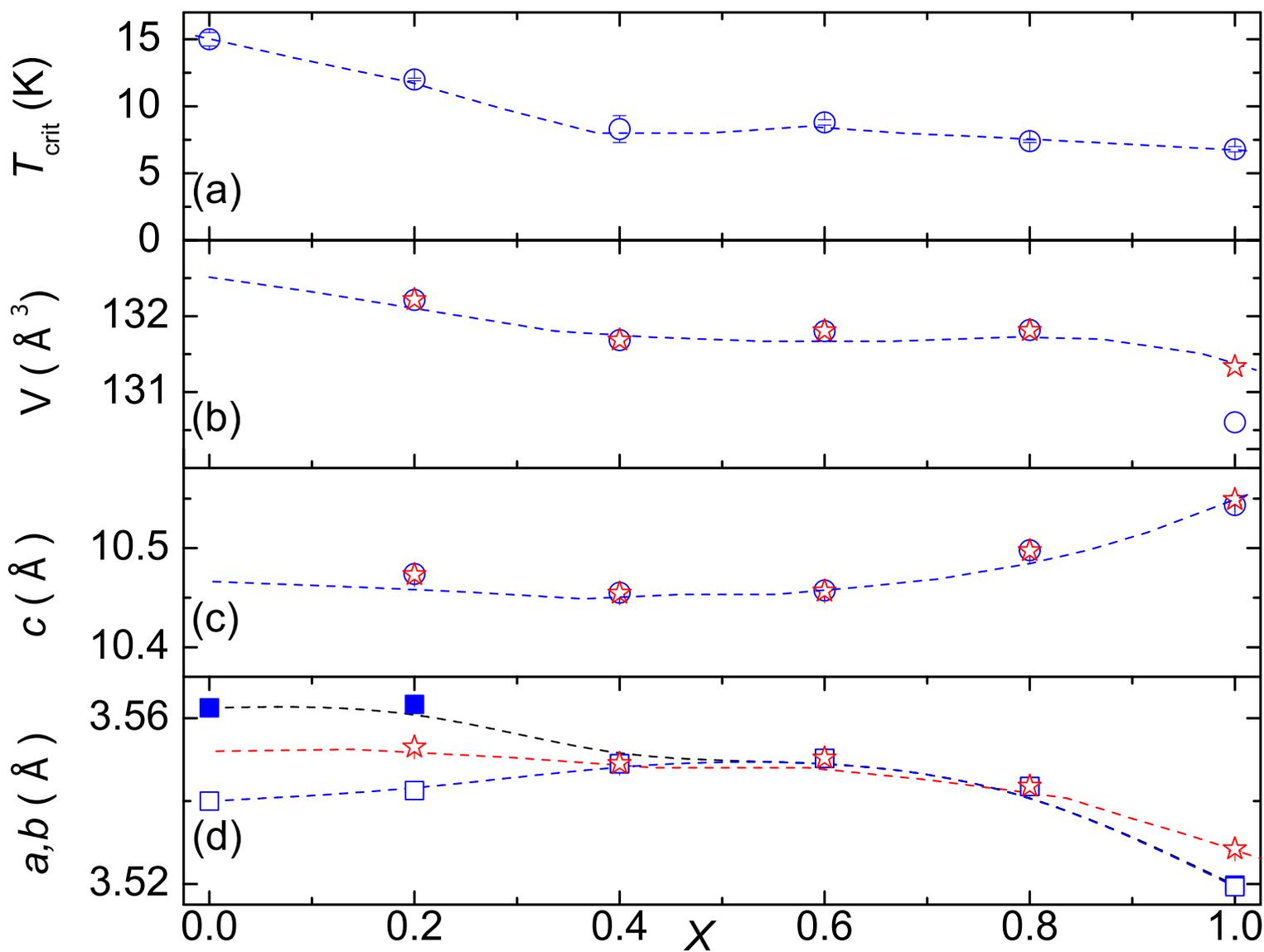
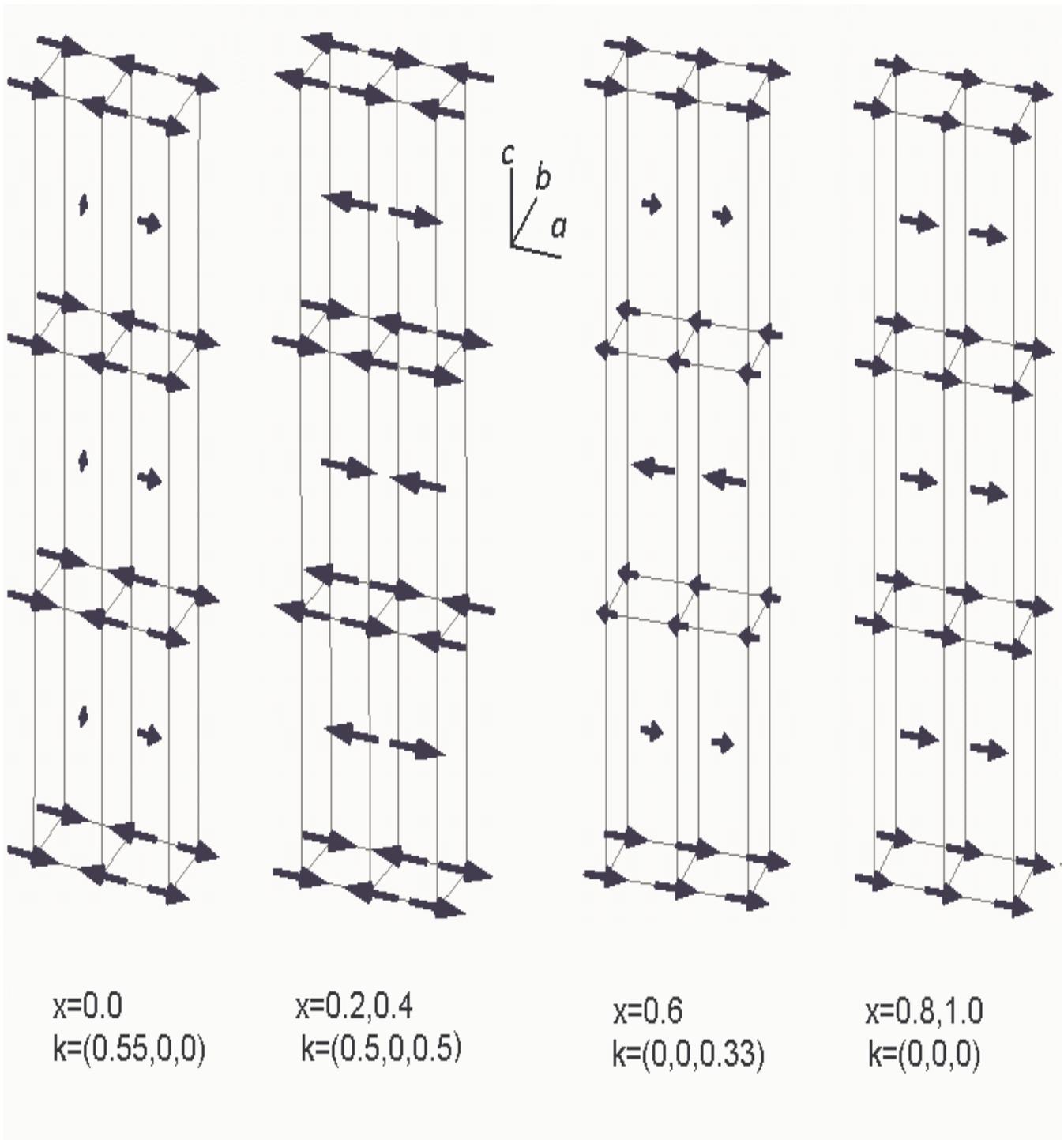
\end{document}